\documentclass[logo,11pt,a4paper]{ETHpaper}
\usepackage{graphicx, amsmath, amssymb,color,wasysym}
\usepackage{bm}
\usepackage[square,numbers,sort&compress]{natbib}
\hypersetup{
     colorlinks=false, allbordercolors={0 0 1}, pdfborderstyle={/S/U/W 1}
    }

\usepackage{caption}
\usepackage{subfigure}
\usepackage{longtable}
\usepackage{booktabs}
\usepackage{mathtools}
\usepackage{cancel,cool}

\usepackage{xcolor}

\begin{document}

\newcommand{\mean}[1]{\left\langle #1 \right\rangle} 
\newcommand{\abs}[1]{\left| #1 \right|} 
\newcommand{\ul}[1]{\underline{#1}}
\renewcommand{\epsilon}{\varepsilon} 
\newcommand{\eps}{\varepsilon} 
\renewcommand*{\=}{{\kern0.1em=\kern0.1em}}
\renewcommand*{\-}{{\kern0.1em-\kern0.1em}} 
\newcommand*{\+}{{\kern0.1em+\kern0.1em}}

\newcommand{\RA}{\Rightarrow}
\newcommand{\bbox}[1]{\mbox{\boldmath $#1$}}

\title{A conceptual approach to model co-evolution of urban structures}

\titlealternative{A conceptual approach to model co-evolution of urban structures}

\author{Frank Schweitzer,\footnote{Corresponding author: \url{fschweitzer@ethz.ch}}
Vahan Nanumyan
}

\authoralternative{F. Schweitzer, V. Nanumyan}

\address{Chair of Systems Design, ETH Zurich, Weinbergstrasse 58, 8092 Zurich, Switzerland}

\reference{Preprint, submitted to \emph{International Journal of Space Structures} (2015)}

\www{\url{http://www.sg.ethz.ch}}

\makeframing
\maketitle

\begin{abstract}
Urban structures encompass \emph{settlements}, characterized by the spatial distribution of built-up areas, but also \emph{transportation} structures, to connect these built-up areas. 
These two structures are very different in their origin and function, fulfilling complementary needs: (i) to \emph{access} space, and (ii) to \emph{occupy} space. 
Their evolution cannot be understood by looking at the dynamics of urban aggregations and transportation systems separately.
Instead, existing built-up areas feed back on the further development of transportation structures, and the availability of the latter feeds back on the future growth of urban aggregations. 
To model this \emph{co-evolution}, we propose an agent-based approach that builds on existing agent-based models for the evolution of trail systems and of urban settlements. 
The key element in these separate approaches is a generalized communication of agents by means of an adaptive landscape.
This landscape is only generated by the agents, but once it exists, it feeds back on their further actions. 
The emerging trail system or urban aggregation results as a self-organized structure from these collective interactions. 
In our co-evolutionary approach, we couple these two separate models by means of \emph{meta-agents} that represent humans with their different demands for housing and mobility. 
We characterize our approach as a statistical ensemble approach, which allows to capture the potential of urban evolution in a bottom-up manner, but can be validated against empirical observations. 

\end{abstract}

\section{Introduction}
\label{sec:Introduction}

The legendary collaborative research project (German abbreviation: SFB) No. 230 ``Natural Constructions'' was established in Stuttgart, Germany, in 1984, the same year the famous Santa Fe Institute (SFI) was established in Santa Fe, New Mexico.  
In the opening workshop of the latter, one of the founders of the SFI, the Nobel laureate \textsc{Murray Gell-Mann}, said:  
``A new subject is taking shape, which has roots in cognitive science, in nonlinear systems dynamics,
and in many parts of the physical, biological, and even the behavioral sciences. Some people call it
self-organization, others complex systems theory, others synergetics, and so forth. It tries to attack the
interesting question of how complexity arises from the association of simple elements.'' \citep{gell-mann}. 

\emph{Complexity} and \emph{interdisciplinarity} are not only the key words to characterize the scientific profile of the SFI, they also describe the  scientific aspiration of the SFB 230 \cite{teichmann96}.
\textsc{Frei Otto} was the \emph{spiritus rector} of this SFB and one of its leading figures during the first eight years.
It was evident to him that the design of urban structures, from the architecture of buildings to the transportation infrastructure of cities and the regional planning of settlements, cannot be understood, and not be revived, without understanding the fundamental principles of self-organization. 
And this cannot be achieved without involving disciplines other than architecture, construction engineering and town planning, i.e. natural sciences such as biology and physics. 

This is the reason why  one of the authors (FS) joined the SFB 230 in early 1992, to contribute to a subproject E2 \emph{Principles of Self-Organization and Evolution} lead by \textsc{Werner Ebeling}. 
Our task was precisely to develop formal models, to explain and to simulate the evolution of urban structures, bottom-up. 
We could build on the phenomenological understanding of these processes already developed by architects and town planners, such as \textsc{Frei Otto} \citep{otto91}, \textsc{Eda Schaur} \citep{schaur-91} and \textsc{Klaus Humpert} \citep{humpert1994phanomen}, just to name a few. 

In November 1991 \textsc{Frei Otto} has just published a small booklet in the \emph{Concept Series} of the SFB 230, a series aimed at steering the discussion rather than publishing firm results.  
It was titled ``The natural construction of grown settlements'' (published in German: Die nat\"urliche Konstruktion gewachsener Siedlungen) \citep{otto91}.
This booklet, in some sense, became the guideline for our research for the coming four years.
It sketched, with the hand written text corrections and hand drawn illustrations by \textsc{Frei Otto}, the two fundamental processes we should model by means of an abstract approach: \textbf{``Erschliessen''}, the process of accessing space, and \textbf{``Besetzen''}, the process of occupying space. 
The paragon for accessing space was the trail system, not just of humans, but also of other biological species (p. 65 of the mentioned booklet contains the trail system of a mice settlement near Warmbronn, hand-drawn by \textsc{Frei Otto}). 
Paragons for occupying space were natural forms such as foams or bubble floats, meshes, but also non-planned human settlements, which are captured in the eminent book by \textsc{Eda Schaur} \citep{schaur-91}. 

\begin{figure}[htbp]
  \centering
  \includegraphics[width=0.5\textwidth]{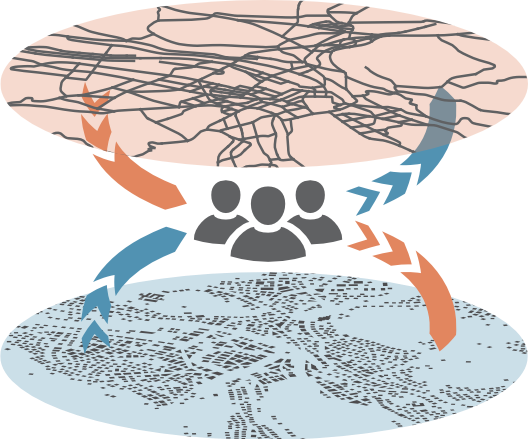}
\caption[]{Two-layer description of urban structures: (bottom layer) urban aggregation, (top layer) transportation system. 
Both layers influence each other in their evolution. The feedback is mediated by some \emph{meta-agents} (see Sect. 3).}
\label{fig:1}
\end{figure}
It was obvious already at the phenomenological level that these two processes of \emph{access} and \emph{occupation}, or \emph{transportation} and \emph{aggregation}, as we will call them in the following, are inherently tight to each other. 
Accessing space is the precondition of its subsequent occupation, but existing occupations also shape the further evolution of the structures that connect them. 
Hence, we face the problem of \emph{co-evolution}, where two levels of different structure and function feed back on each other (see Fig. \ref{fig:1}). 
This clearly defines the problem we need to solve: (1) to model, for each layer \emph{separately}, the evolution of the structure, e.g. the trail system and the urban aggregation, and (2) to combine these two layers in a generalized system model, to study their mutual feedback and co-evolution.

This sets the stage for the rest of this paper. 
We will first discuss the general concept for modeling these structures by means of an adaptive landscape.
Then, we demonstrate by means of examples how such structures evolve, for  transportation and aggregation separately.
Eventually, we sketch how a model to combine these two layers shall look like.

\section{Agent-based models of urban structures }
\label{sec:agent-based-models}

\subsection{Generalized communication}
\label{sec:gener-comm}

Methodologically, we follow the bottom-up approach, i.e. we start from the mentioned question of how ``complexity arises from the association of simple elements.'' 
These elements, commonly denoted as \emph{agents}, represent the units of the system which generate the structure. 
Agents are a rather abstract representation of entities with a certain \emph{demand}. 
In line with the problem description given above, we use two different types of agents, entities with the need of \emph{assessing space} and entities with the need of \emph{occupying space}. 
These agents follow a given dynamics, i.e. the need transforms into some sort of \emph{activity} in time, which is in our case to move (accessing space) and to aggregate (occupying space).
That means, agents are not simply equal to e.g. \emph{humans}, although humans and other biological species assess and occupy space.

Urban systems can be seen as instances of \emph{complex systems}, i.e. they consist of a large number of heterogeneous agents that are ``similar'', but not identical with respect to their properties. 
The \emph{interaction} between these agents on the \emph{micro level} results in the formation of urban structures on the \emph{macro level}. 
This is often denoted as \emph{emergence},  the sudden occurence of \emph{new system qualities} once certain critical parameters, known as thresholds or tipping points, are reached.
These new system qualities cannot be decomposed or reduced to the properties of individual agents, which is a feature of all self-organizing systems. 
\emph{Self-organization} describes ``the process by which individual subunits achieve, through
    their cooperative interactions, states characterized by new, emergent properties transcending
    the properties of their constitutive parts.''\citep{biebrep-95}

How shall we then model the ``cooperative interactions'' between a large number of agents, in a general way?
Today, the \emph{complex network approach} has become fashionable. 
It decomposes all interactions between agents into \emph{binary} interactions, i.e. interactions  between \emph{two individual} agents which are represented by links, while the agents are represented by the nodes of the network. 
Such a description has many disadvantages if we want to model urban structures.
First of all, we have to consider the (two-dimensional) \emph{physical space}, i.e. interactions between agents are bound to some defined \emph{spatial neighborhood}.
Secondly, in many situations agents do not interact directly, but \emph{indirectly} by means of a medium.  
Taking the example of an emerging trail system, agents are not attracted to other agents but to the trail they commonly use. 

\begin{figure}[htbp]
  \includegraphics[width=0.45\textwidth]{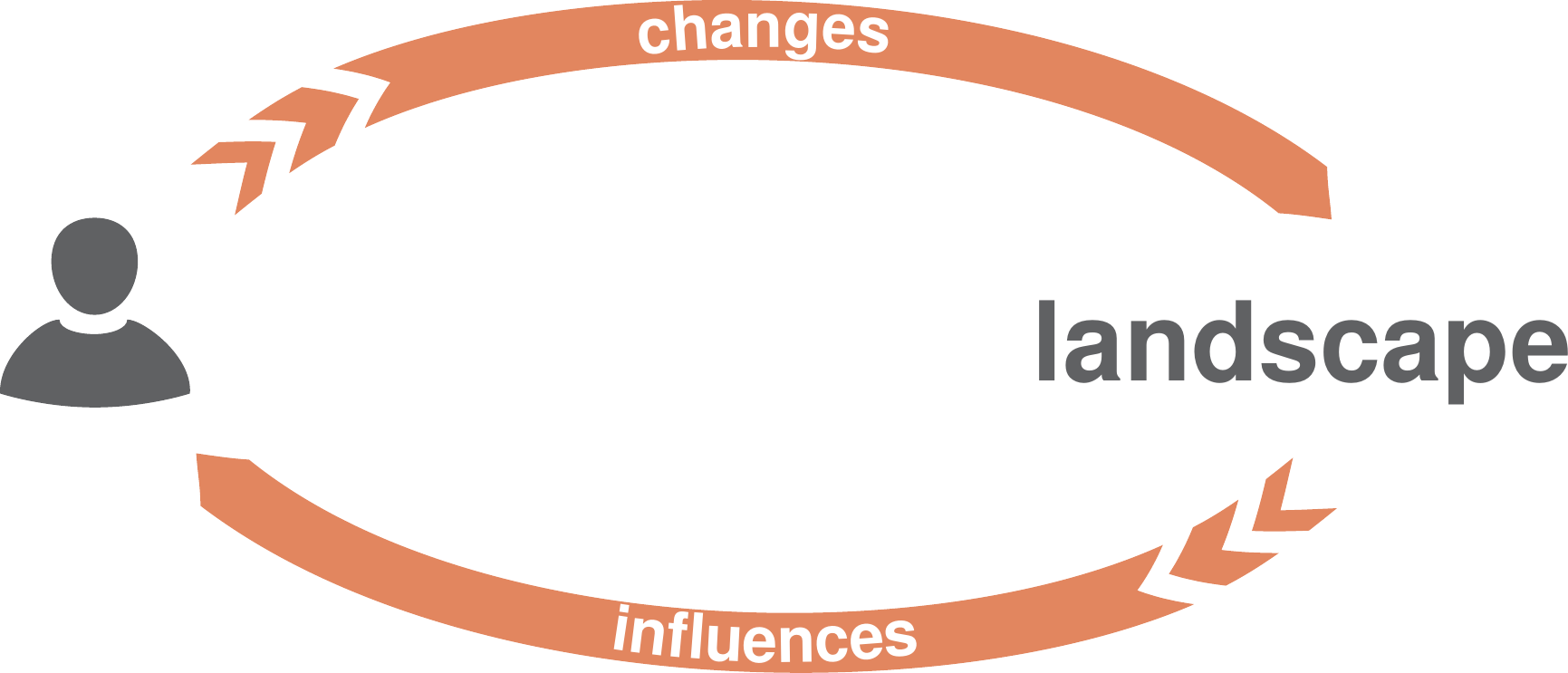}\hfill
  \includegraphics[width=0.45\textwidth]{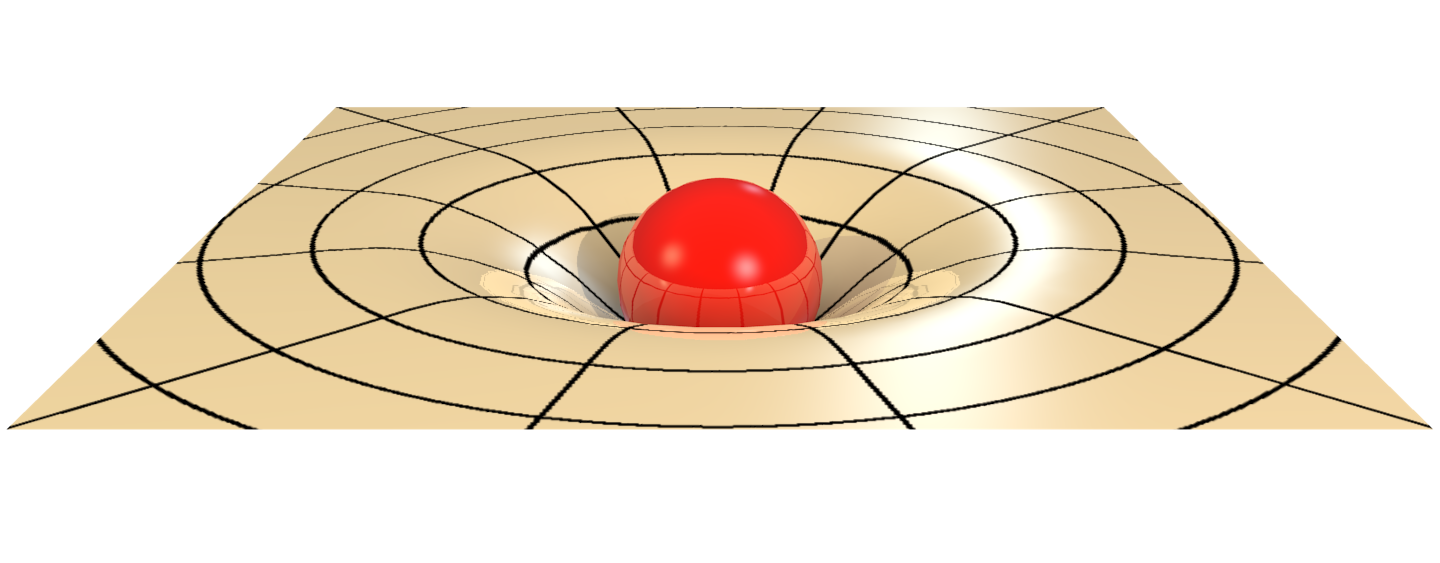}
  \caption{(left) Feedback between agents and the adaptive landscape, (right) mass curves space which influences other masses.}
  \label{fig:feedback}
\end{figure}

We can describe this kind of interaction as generalized \emph{communication} \citep{fs-97-wf}, i.e. agents ``read'' and ``write'' information which is exchanged by a ``communication field''. 
The latter serves as a medium that couples the different agents in a weighted manner, i.e. it takes the spatial distance, the dissemination and the aging of information into account. 
With respect to urban structure formation, we can see this communication field as an \emph{adaptive landscape} that is shaped by the actions of all agents collectively, but also feeds back on their actions. 
This feedback is described in Fig. \ref{fig:feedback}. 

We will illustrate the role of the adaptive landscape and its meaning in urban structure formation in the following sections.
But before, we want to make a general comment.
At a time, where we celebrate the 100th anniversary of \textsc{Albert Einstein}'s \emph{general relativity theory}, it is worth noticing that even in physics particles ``communicate'' indirectly via fields. 
Electrons generate an electrostatic field that ``communicates'' their position and electric charge to other particles. 
And these particles ``respond'' differently to this information. Positrons are attracted, whereas other electrons are repelled. 
In the same vein, mass generates a gravitational field. 
More precisely, as \textsc{Albert Einstein} noticed in his seminal theory, mass \emph{curves} physical space which in turn influences the motion of other masses and even of light. 
So, physical space can be seen as an adaptive ``landscape'' that constantly adapts to the distribution of mass while affecting its position. 

In the following, we build our conceptual approach for the evolution of urban structures on such \emph{adaptive landscapes}. 
These landscapes are only generated by the agents and in turn influence their further action. 
But they can also follow an \emph{eigendynamics}, i.e. the information contained in these landscapes can diffuse and decay by itself, without the involvement of agents.

\subsection{Agent-based models of trail formation}
\label{sec:agent-based-models-1}

Trail formation gives a lucid example of how the adaptive landscape is generated by the agents.
We assume that agents move in a two-dimensional physical space and leave a marker at each position they visit (``writing'').
These markers can be sensed by other agents if they are in the immediate vicinity (``reading'').
Agents then decide with a certain probability to follow the existing markers (``acting'').

Ants, for example, use different chemical markers, so-called pheromones, to mark their trails and to provide cues for other ants.
All the markers together define the information field $h(\mathbf{r},t)$ that depends on the position $\mathbf{r}$ in the two-dimensional space and on time $t$.
This field follows its own dynamics: if no new markers are created at a given position, the field decays over time, e.g. the chemicals decompose. 
This way, the field constantly adapts to the current movement of the agents, it increases at positions highly visited and decreases otherwise.

\begin{figure}[htbp]
  \centering
  \includegraphics[width=0.5\textwidth]{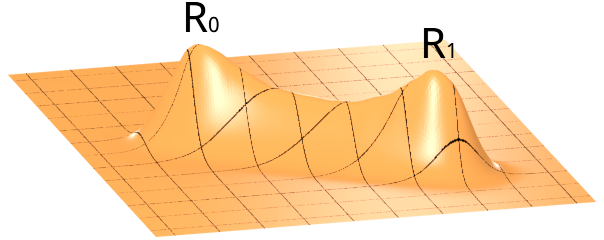}\caption{An adaptive landscape representing a trail between two locations $R_{0}$, $R_{1}$ \citep{Perony2012}.}
  \label{fig:gradient}
\end{figure}
Agents evaluate this information by measuring the \emph{gradient} of the field, i.e. they prefer to move into the direction of \emph{higher} values.
Their motion can be seen as a hike in the adaptive landscape that is changed by every step.
Agents try to follow a route along the ``mountain ridge'' (see Fig. \ref{fig:gradient}).
This reinforces the existing markers, which in turn attracts more agents. 
Eventually, all agents move along the same trail which becomes visible by the high concentration of information, e.g. of chemical markers. 
Pedestrians may not use such markers, but they leave footprints cutting the grass which serves the same purpose \citep{helbing-fs-et-97}. 
\begin{figure}[htbp]
  \includegraphics[width=0.45\textwidth]{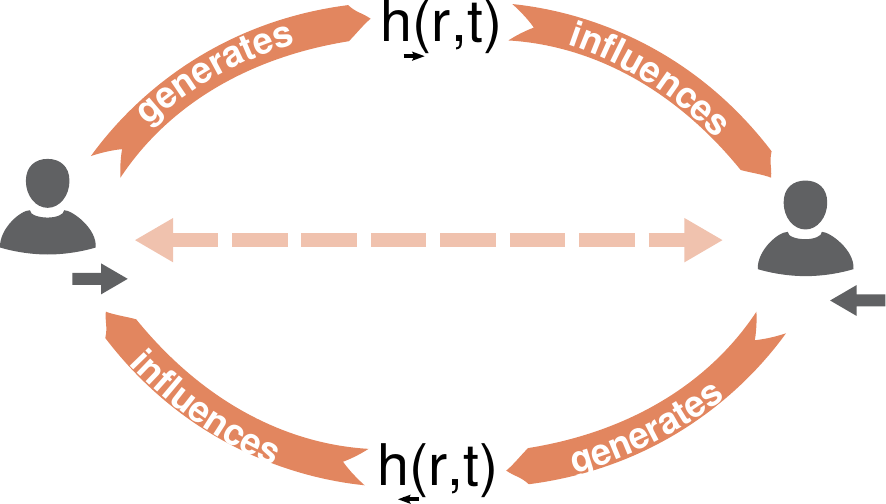}\hfill
  \includegraphics[width=0.30\textwidth]{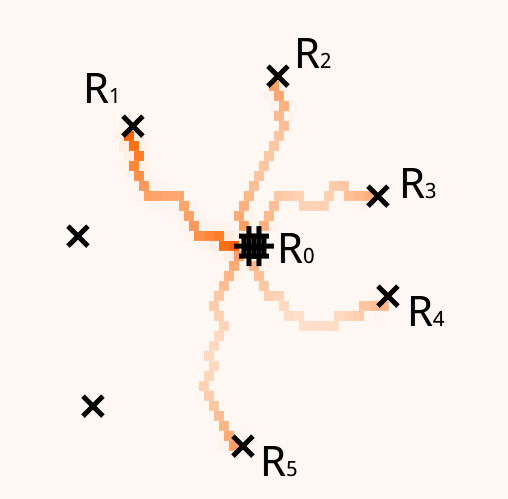}\hfill
  \caption{(left) Feedback between agents and the adaptive landscape in case of directed trails. (right) Self-organized trails to connect a center with locations at the periphery \citep{fs-tilch-02-a}}
  \label{fig:trail}
\end{figure}

To observe directed movements between two locations $A$, in the center, and and $B$ in the periphery, 
we use two different kind of markers that generate their own field. 
In the example of Fig. \ref{fig:trail}, we note that the information \emph{generated} by the agents moving from $A$ to $B$ 
affects agents moving into the opposite direction, and vice versa. 
This indirect feedback helps agents starting from $A$ to find their way to any of the peripheral locations $B$ and the other way round. 
If agents would only follow their own information, they would get stuck in either $A$ or $B$ 
because the gradient would always point back to their origin.

\subsection{Agent-based models of urban aggregation}
\label{sec:agent-based-models-2}

Similar to trail systems, also the evolution of urban aggregates can be modeled by means of agents creating, and interacting with, an adaptive landscape.
We note that in this case the agents, in an abstract manner, represent entities with a different demand, namely of \emph{occupying space}, which also translates into a different \emph{activity}, namely to aggregate.
Occupying space depends on two kind of ``resources'', a \emph{demand} (represented by the agent) to occupy a (free) site  and a \emph{supply}, i.e. the availability of free sites.
Hence, we face the problem to first \emph{match supply and demand}, which only leads to the desired activity, to aggregate. 
Therefore, occupation combines two \emph{different} processes, the \emph{search process}, to find the right place, and the process to \emph{settle}, i.e. a transition from being mobile during the search to becoming immobile.     

\begin{figure}[htbp]
  \includegraphics[width=0.5\textwidth]{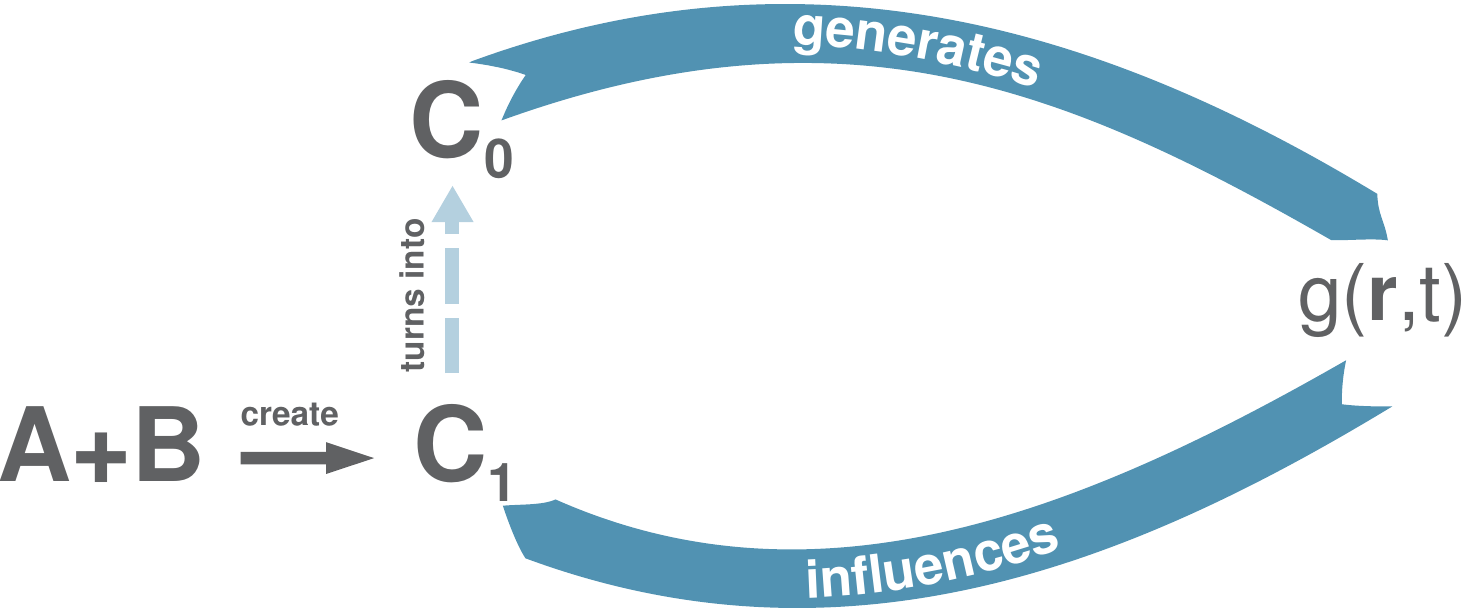}\hfill
  \includegraphics[width=0.37\textwidth]{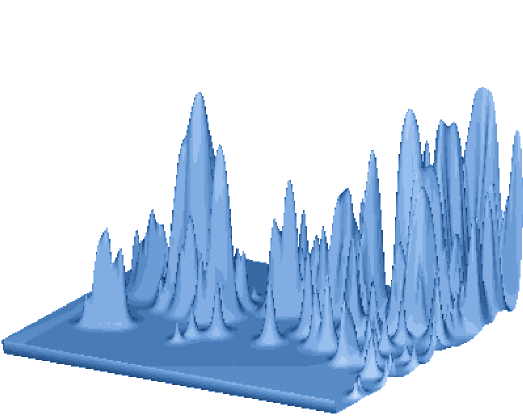}\hfill
  \caption{(left) Feedback between agents and the adaptive landscape for urban aggregation. (right) Urban attraction field of the south-east built-up area of Berlin/Potsdam (1910) \citep{fs-steinb-97-gb}}
  \label{fig:urban}
\end{figure}
Both the search and the settle processes depend on information about the already existing urban aggregation, which is captured in  
an adaptive landscape.
While this information is generated by the existing build-up area, it can also spread out to the neighborhood.
For example, downtown Manhattan creates an attraction potential that also spills over to adjacent areas. 
To account for this, the adaptive landscape, denoted by $g(\mathbf{r},t)$ for urban settlements, is created by agents that represent \emph{existing built-up units} $C_{0}$, which do not move (see Fig. \ref{fig:urban}). 
Another type of agents, $C_{1}$, represents \emph{growth units}, which are \emph{potential} build-up units searching for the best location to ``settle''. 
These units are created far outside of the center, taking into account \emph{free space}, $A$, and \emph{demand} for settlement space, $B$.  
In their search for an optimal location, the $C_{1}$ agents follow the gradient of the adaptive landscape, trying to get as close as possible to the maximum. 
The rate at which growth units transform into real build-up units, i.e. become immobile, also increases with the value of $g(\mathbf{r},t)$. 
That means the $C_{1}$ agents likely ``settle'' before they have reached the maximum of the adaptive landscape. 

This agent-based model, known as $A$-$B$-$C$ model \citep{fs-steinb-97-gb}, is able to reproduce two empirical observations in the growth of large urban settlements: (a) as time progresses, urban growth zones shift toward outer regions of the urban area and sometimes concentrate around suburbs, (b) the urban center, although most attractive, does not further grow in size.  
The latter means that construction activities aim at reusing existing built-up areas, but not to fill existing free space. 
The fractal-like structure of and the cluster size distribution \citep{fs-steinb-98-app} of urban settlement areas is still kept. 

\begin{figure}[htbp]
  \centering
  \includegraphics[width=0.30\textwidth]{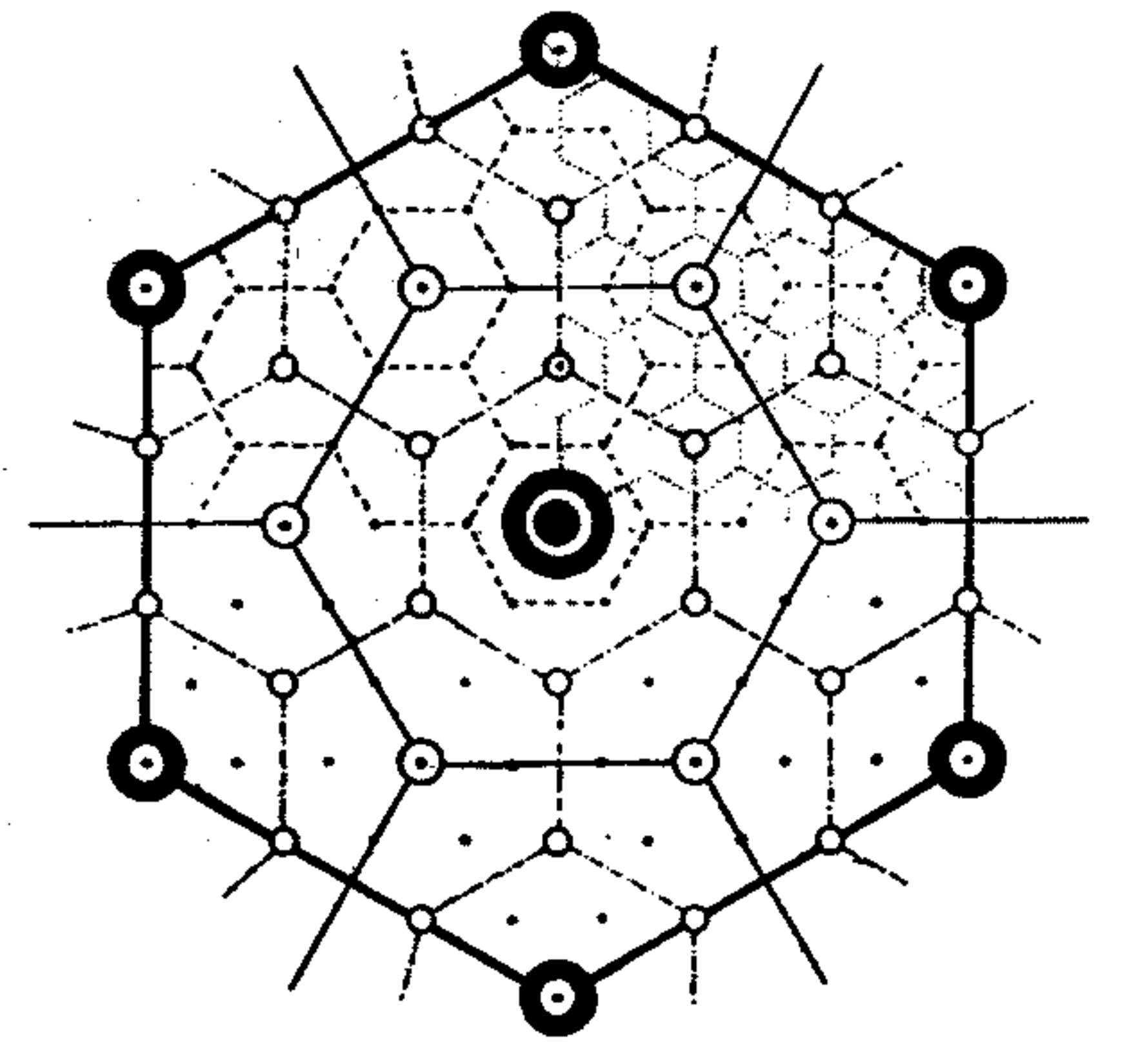}\hfill
\includegraphics[width=0.65\textwidth,height=10em,angle=0]{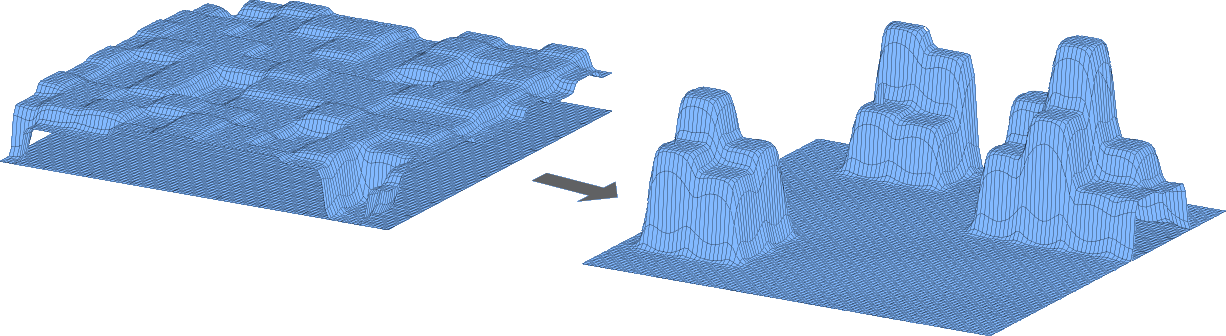}
  \caption{(left) Sketch of the spatial distribution of central places of hierarchical importance \citep{christaller-33}. (right) Emergence of economic centers from a broad spatial distribution economic activities \citep{fs-98-jcs}.}
  \label{fig:product}
\end{figure}
We note that a similar agent-based model \citep{fs-98-jcs} is able to reproduce the emergence of urban centers at a certain critical distance from each other.
This was one of the key propositions of the \emph{central place theory} developed by \textsc{Walter Christaller} \citep{christaller-33} in 1933 (see Fig. \ref{fig:product})
In this model, the adaptive landscape reflects the spatial distribution of \emph{production}, which defines an average wage paid to the workers at a particular location. 
Agents represent either employed ($C_{0}$) or unemployed ($C_{1}$) workers. 
Unemployed workers can migrate and take into account \emph{gradients} in the wage, i.e. they prefer to move to places that pay a higher wage. 
If they become employed workers, they no longer migrate but settle and start contributing to the production, this way increasing the average wage at that place. Figure \ref{fig:product} shows such a spatial distribution of production centers from a simulation. 
The critical distance between the centers allows them to \emph{coexist}, in agreement with the theory of \textsc{Walter Christaller}. 
But the emergence of the critical distance is a self-organized phenomenon, resulting from the transition of broadly distributed centers with low productivity to localized distant centers with high productivity.

\section{Co-evolution of urban structures}
\label{sec:co-evolution-urban}

So far, we have outlined a conceptual approach to model trail systems and urban aggregation, \emph{separately}. 
We now want to combine these two subsystems into a model of \emph{co-evolution}.
As already explained, trail systems, or transportation systems in general, allow to \emph{access} space, which is the precondition of urban settlements.
But existing urban aggregations also shape the way the transportation system evolves further, i.e. there is a  mutual feedback between \emph{transportation} and \emph{aggregation} as indicated in Fig. \ref{fig:1}. 

We note that, within our conceptual approach, each of these subsystems is described by an \emph{adaptive landscape}, which is generated by the agents and feeds back on their further options, to move or to settle. 
So, it is natural to assume that the co-evolution is modeled by \emph{combining} these two adaptive landscapes. 

\begin{figure}[htbp]
\centering
    \includegraphics[width=0.75\textwidth]{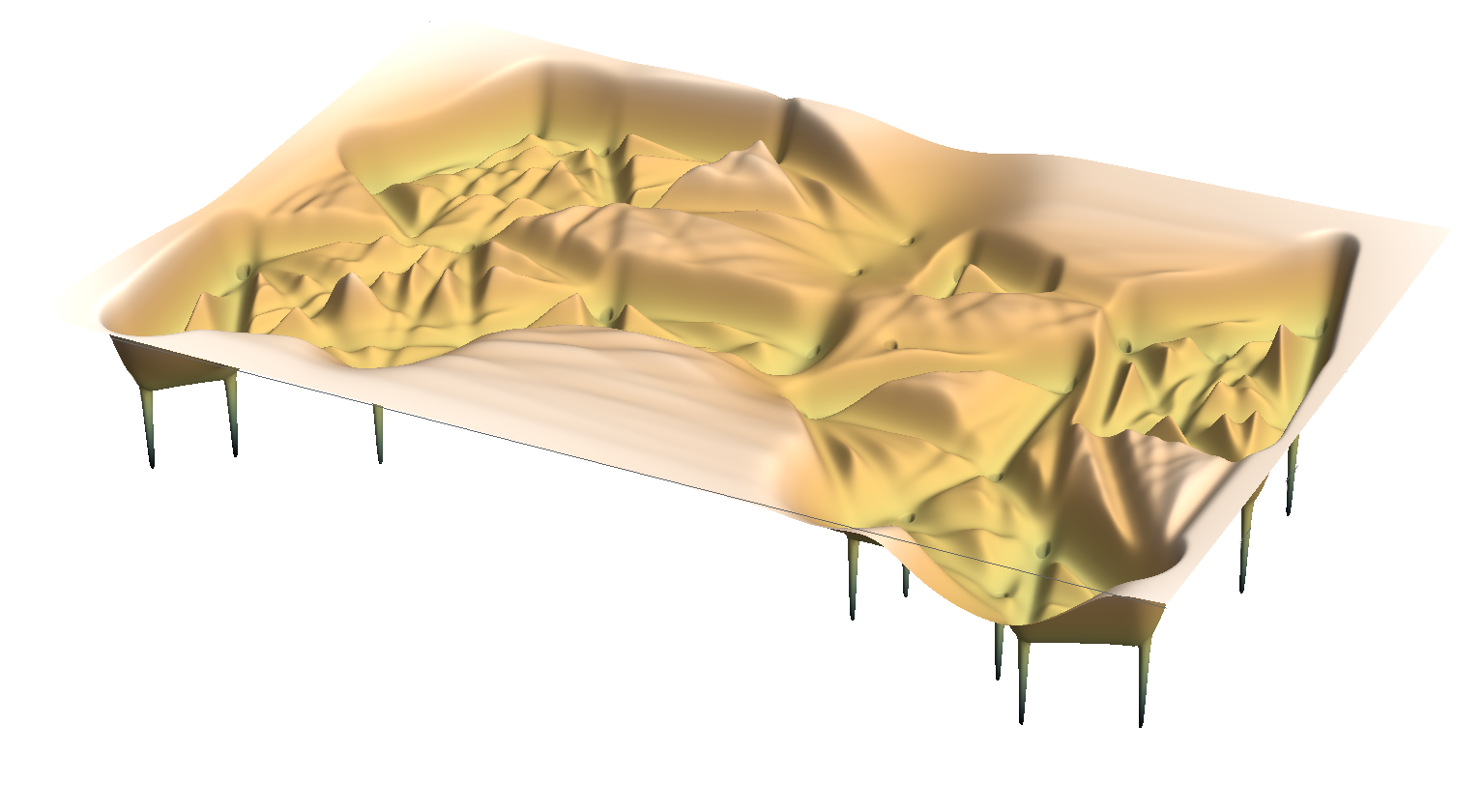}
    \caption{Adaptive landscape of a mice settlement with trails and nest boxes \citep{Perony2012}.}
\label{fig:mice}
\end{figure}
Figure \ref{fig:mice} gives an example of such a combined landscape for a mice settlement \citep{Perony2012}.
These mice live inside a barn (with the option to enter and leave), where they also find food. 
They shelter in different nestboxes and can freely move between them. 
These boxes represent the built-up areas that are attractive to the mice. 
There attractiveness is indicated in the adaptive landscape by narrow spikes that point down, i.e. mice try to move to the \emph{minima} of the landscape (because here the landscape is inverted for a better view). 
The movement of the mice occurs along preferred routes in the barn that can be identified in the adaptive landscape as valleys with a straight orientation. 
The deeper the valleys, the more frequently they are used. 
Areas with higher elevation indicate that mice do not move there and also do not settle there. 

While this picture gives us a graphic idea of how such a combined landscape, covering settlements and trail systems, shall look like, it has the major drawback of being \emph{static} instead of \emph{dynamic}. 
The locations of the nestboxes are given and the spikes are therefore imposed to the landscape, but at least the trail system in reality constantly adapts with respect to the usage by the mice. 

\begin{figure}[htbp]
\centering
\begin{minipage}[c]{0.5\linewidth}
  \includegraphics[width=0.99\textwidth]{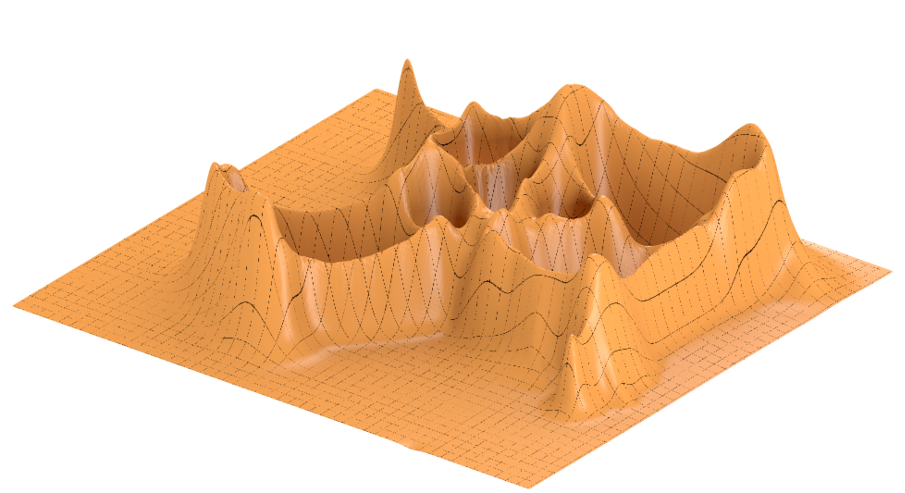}\\
    \includegraphics[width=0.99\textwidth]{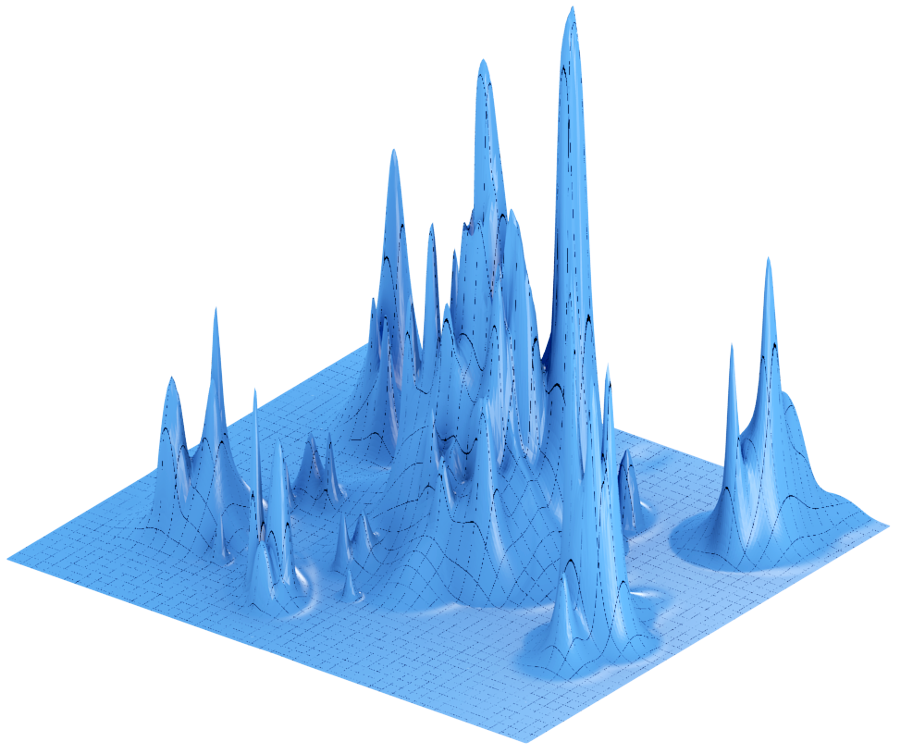}
\end{minipage}\hfill
\begin{minipage}[c]{0.5\linewidth}
\includegraphics[width=0.99\textwidth]{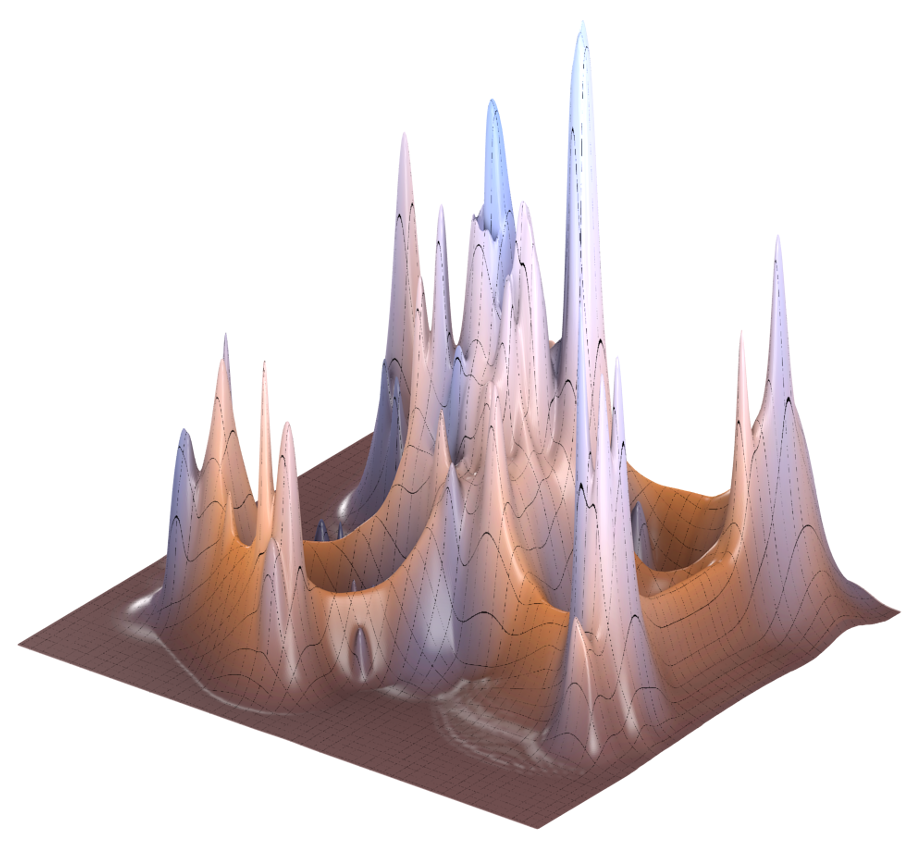}\\
\vspace*{-6.5cm}  
{\includegraphics[width=0.2\textwidth]{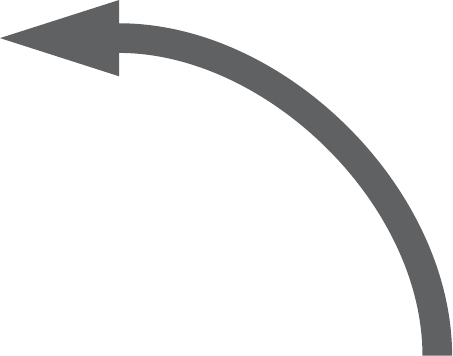}}\\
\vspace*{4.5cm}
\reflectbox{\rotatebox[origin=c]{180}{
\includegraphics[width=0.2\textwidth]{arrow}}}
\end{minipage}
\caption{Conceptual representation of the combined adaptive landscape (mixed color) and its decomposition into the two adaptive landscapes for transportation (orange) and urban aggregation (blue).}
\label{fig:superposition}
\end{figure}

To capture the \emph{co-evolution} between transportation and aggregation structures, we utilize a \emph{two-layer approach}, also shown in Fig. \ref{fig:1}. 
Each layer contains only one structure, either aggregation or transportation. 
To obtain the same for the information field, we need to disentangle the combined adaptive landscape, as it is shown in Fig. \ref{fig:superposition}.  
Both layers now contain \emph{different information}, either about the existing transportation structure, $h(\mathbf{r},t)$, or about the existing urban aggregation, $g(\mathbf{r},t)$. 
But they use the same \emph{spatial coordinate system}, i.e. a specific location is represented in both layers at the same position $\mathbf{r}$.

The main challenge is then to model the feedback between these two layers.  
This is indicated in Fig. \ref{fig:1} by means of some ``agents''.
These are obviously not identical to the rather abstract agents representing either the need of accessing or occupying space, in each layer separately.  
Instead, these are \emph{meta-agents} which \emph{combine} these different needs, and they can be seen more like \emph{humans}. 
To elucidate how such a combination could look like on a mathematical level, let us assume that agents on the \emph{transportation layer} are described by a function $\mathcal{F}[x,\mathbf{u},h(\mathbf{r},t)]$, where $x$ denotes the agent, $h(\mathbf{r},t)$ the information field that couples the  processes to collectively generate the transportation structure and $\mathbf{u}$ a set of control parameters to represent the boundary conditions. 
Agents on the \emph{occupation layer}, on the other hand,  are described by a function $\mathcal{K}[y,\mathbf{v},g(\mathbf{r},t)]$, where $y$ denotes the agent, $g(\mathbf{r},t)$ the information field that couples the  processes to collectively generate the urban aggregation and $\mathbf{v}$ a set of control parameters to represent the respective boundary conditions. 
The \emph{meta-agent} $z$ then is described by a function $\Omega\{z,\mathcal{F},\mathcal{K}\}$ that combines these two needs and, hence, considers the two different kind of information stored in the adaptive landscapes $h(\mathbf{r},t)$ and $g(\mathbf{r},t)$.
What sounds rather abstract is a mathematically convenient, and transparent, way of combining the separate models. 
But we are not discussing specific forms for the functions $\Omega$, $\mathcal{F}$ and $\mathcal{G}$ here.

\begin{figure}[htbp]
\centering
    \includegraphics[width=0.9\textwidth]{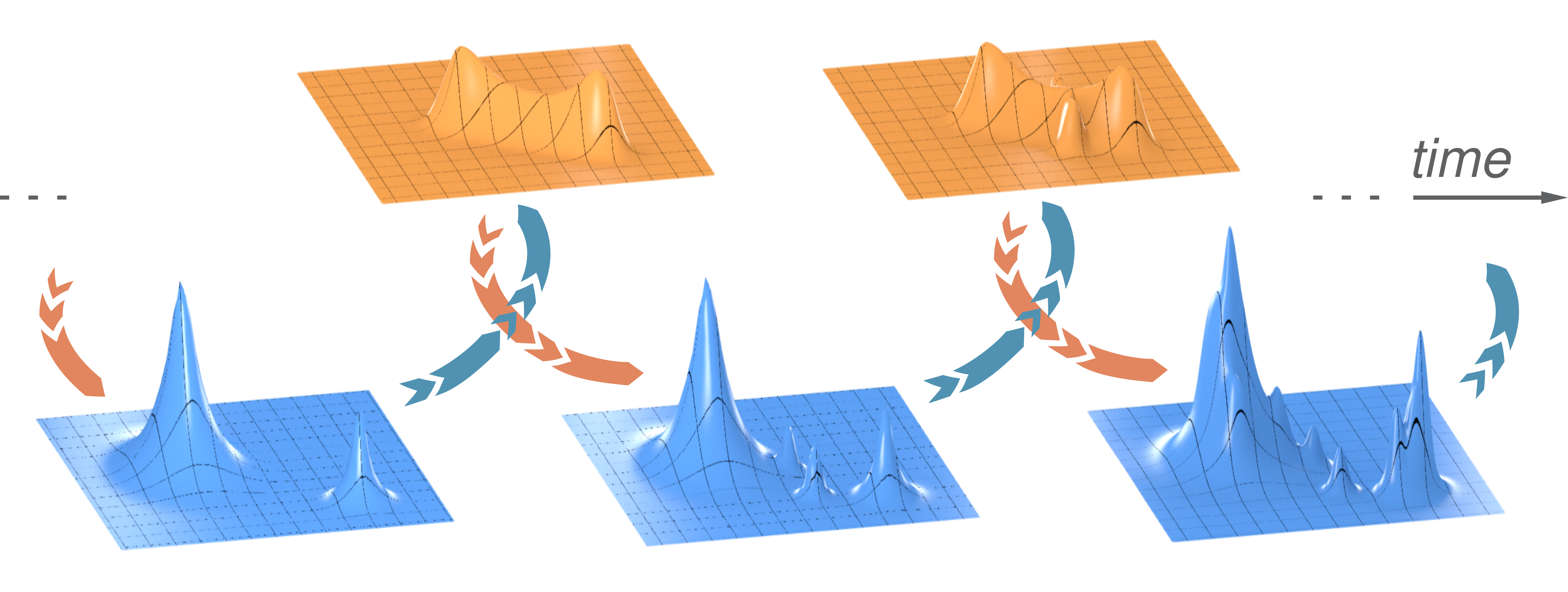}
    \caption{\emph{Co-evolution} of two different urban structures, aggregations (lower layer) and transportation structures (upper layer), illustrated by means of their respective adaptive landscape. Arrows indicate the feedback between the two layers.}
\label{fig:evo-steps}
\end{figure}

Figure \ref{fig:evo-steps} illustrates the \emph{co-evolution} of the transportation and the aggregation layers by 
showing the respective adaptive landscapes.
Changes in the urban aggregation feed back, at a later time step, on the transportation structure via the \emph{meta-agents} that adjust their \emph{demand} for transportation based on the recent \emph{supply} of built-up area. 
The resulting changes in the transportation structure in turn feed at the next time step on the aggregation structure. 
Precisely, a \emph{supply} in transportation at a given time results in new attraction zones for the growth units that create the built-up area afterwards.
If new build-up areas are supplied, this generates a new demand for transportation and so forth. 
Hence, both layers co-evolve in time. 
 
\section{Conclusions}

In the following, we move the previous discussion to a more general level, addressing some pros and cons of our conceptual approach. 

\paragraph{System representation}

There are different modeling approaches for the dynamics of systems comprised of a large number of interacting elements.
The so-called \emph{systems dynamics} approach builds on \emph{representative agents}, i.e. instead of many similar agents interacting \emph{one typical} agent is used to represent all agents of that kind. 
A prominent example is \emph{macro economics}, where models use a small number of different agents, e.g. \emph{the} firm, \emph{the} customer etc., to focus on the \emph{nonlinear feedback} between these representative agents.

The \emph{multi-agent} approach, on the other hand, builds on the interaction between a large number of individual agents and the focus is on the emergent system properties, not on the role of  single agents. 
These agents are \emph{heterogeneous}, i.e. they are similar, but not identical, and there can be different types of agents in the model. 

To model the emergence of urban structures, we have used two different types of agents, one representing the need of \emph{assessing space}, the other one representing the need of \emph{occupying space}. 
These agents are a rather abstract representation of a certain \emph{demand} that has to be satisfied \emph{collectively}. 
To model the co-evolution of urban structures, we combine these different needs in \emph{meta-agents} that can be seen more like humans. 

\paragraph{Bottom-up approach}

Agent-based modeling is essentially a bottom-up approach, which means that in our model there is no hierarchical planning or centralized control of the processes generating urban structures. 
Instead, these structures \emph{emerge} from the collective interaction once critical tipping points are crossed.
As with all self-organizing processes, it remains a challenge to predict \emph{when} this is the case and \emph{how} these structures eventually will look like. 

This raises the question how such processes can still be influenced.
As any other processes, self-organization depends on \emph{boundary conditions} that set limits e.g. to the urban structures that can potentially emerge. 
In our case, these boundary conditions are given by the physical and political geography of the area (lakes, deserts, borders), the topology of the landscape (mountains and valleys), but also by available \emph{resources}, e.g. by the free space that can be potentially accessed/occupied.
Hence, it is possible to \emph{design} (some) boundary conditions, e.g. by restricting the access to space or by limiting resources for transportation.
These conditions then \emph{limit} the possible urban structures, but do not explain which of these emerge. 

We can also influence the interaction between agents, for example their contribution to or their response to the information field generated collectively. 
If the attraction of existing urban structures is increased, this will lead to denser occupation patterns and more concentrated transportation structures. 
Hence, while our modeling approach does not lead to pre-determined structures, it still allows to vary, and to influence, some of the properties on the ``macroscopic'', or systemic, level by controlling interaction properties on the ``microscopic'', or agent, level.

\paragraph{Statistical ensembles}

Architects and town planners may wish for simulation tools that generate life-like visualizations of urban processes. 
This is precisely \emph{not} the aim of our conceptual approach to model urban structures. 
Like a flight simulator, such simulation tools can be quite helpful to learn to ``fly'', but they are essentially not useful to \emph{understand} the system, i.e. to identify the driving factors of its dynamics. 

We aim at a \emph{minimalistic modeling approach}, to highlight the \emph{generic features} of a \emph{whole class} of urban structures.
We follow the principle of \textsc{Occam}'s razor, or \emph{lex parsimoniae}, to only consider the minimal set of assumptions needed to explain a certain phenomenon. 
Therefore, our approach does not contain as much detail as \emph{possible}, but only as much as \emph{necessary} to obtain emergent urban structures. 
This helps us to understand what assumptions are essentially \emph{not needed} to make the outcome happen, but are a \emph{nice-to-have} modeling ingredient to produce a more life-like outcome. 
In order to focus on the emergence of systemic properties, it is also important to \emph{not} already encode the expected outcome into the model. 
For example, preferred areas for urban settlements have to be a \emph{result} rather than an input of the model.

How does our approach cope with the mentioned \emph{limited predictability} of urban structures? 
Of course, when we run computer simulations of the agent-based model implemented, we will receive in each run a (slightly) different outcome for the aggregation and transportation patters.
This way, our modeling approach generates a \emph{statistical ensemble} of possible outcomes that are all compatible with the given interactions and boundary conditions.
I.e., it \emph{highlights} the inherent potential for the urban development, instead of focusing on a designed solitary solution. 

Hence, our approach results in a so called \emph{null model} for urban structures that defines a class of possible solutions. 
A null model is a powerful tool for testing statistical hypotheses.
If it is a good null model, then the (one) realized solution will be part of this ensemble. 
But even if it is not, we can get a more fundamental understanding of urban processes by analyzing the \emph{differences} between the modeled structures and the real ones. 
Such deviations then may lead us to the \emph{heart of urban planning}, distinguishing the outcome of generic principles from the impact of \emph{design}, to obtain an \emph{optimized} solution.

\paragraph{Calibration and validation}

How can we then know that our modeling approach is still correct? 
We argue that the model is valid if it is able to reproduce \emph{stylized facts} which are, according to the economist \textsc{Nicholas Kaldor} \citep{kaldor61}  ``stable patterns that emerge from many different
      sources of empirical data, that is, observations made in so many
      contexts that they are widely understood to be empirical truths, to
      which theories must fit.'' 

Such stylized facts are, with respect to economic geography, already summarized in \textsc{Walter Christaller}'s \emph{central place theory} \citep{christaller-33}, pointing out to characteristic distances between urban centers at different levels of hierarchy.   
For the case of non-planned settlements and transportation systems the stylized facts about urban structures are captured in the eminent book by \textsc{Eda Schaur} \citep{schaur-91}. 
For more specific observations, like the fractal structure of urban settlements, books by \textsc{Klaus Humpert} \citep{humpert1994phanomen,becker}, \textsc{Pierre Frankhauser} \citep{frankh-94-book} or \textsc{Michael Batty} \citep{batty-long-94} have contributed to identify stylized facts about the shape, the cluster sizes and the spatial  distribution of built-up areas. 

For our modeling approach, \emph{stylized facts} form the reference point, rather than specific, and often singular, historic observations. 
From these ``robust patterns'', we derive input parameters needed to set-up agent-based computer simulations, such as the mean density of settlements, its fractal dimension, characteristic distances between centers, but also extrapolations for the demand for built-up area. 
We cannot, however, infer from these patters specific model parameters such as the attraction strength of existing aggregations, the decay rate of the information field, the sensitivity toward such information, etc. 
Those model parameters can be only found by comparing the simulation outcome with the stylized observations.
Hence, in all cases, we need a \emph{sensitivity analysis} to estimate the impact of certain model parameters on the aggregated outcome.

\paragraph{A multi-layer approach}

The emphasis of our modeling approach is \emph{not} on simply reproducing settlement patterns or transportation structures, a task already addressed in the stand-alone models. 
Our main focus is on the \emph{co-evolution} of these two urban structures that are very different in their origin and function.
To capture this co-evolution as shown in Fig. \ref{fig:evo-steps}, we utilize a two-layer approach. 
Each layer contains only one structure, either aggregation or transportation, and its dynamics is governed by different kind of agents representing different needs. 
The important idea in our approach is the feedback between these two layers, modeled by \emph{meta-agents} that combine the different agents from each layer. 
This allows to consider the impact of one structure on the other one, e.g. the impact of transportation on the adaptation of the urban settlement pattern. 
This adaptation causes an impact back on the transportation structure to cope with the further demand resulting from the existing urban settlement and so forth. 

The driving force behind this urban co-evolution is the \emph{demand} for new built-up areas, which is essentially driven by the \emph{growth of population} in urban areas. 
This is assumed as \emph{exogenous} to our modeling approach. 
This demand, together with the availability of free space, determines the \emph{growth rate} of the urban settlement (``how much?''), but \emph{not} the \emph{spatial distribution} (``where?'').
The latter depends on the attraction of the existing built-up area, but also on the availability of transportation means, to access space. 
Without existing settlements, there is no demand to expand transportation and without existing transportation, there is no possibility to expand urban settlements. 
Hence, it is essentially not possible to understand, or to model, the change of urban settlement structures without the perspective of \emph{co-evolution}.

\end{document}